\documentclass{article}
\usepackage[utf8]{inputenc}
\usepackage[T1]{fontenc} 
\usepackage{lmodern} 
\usepackage[english]{babel} 
\usepackage{csquotes} 
\usepackage{amsmath,amsfonts,amssymb} 
\usepackage{amsthm} 

\usepackage{tabularx}
\usepackage{booktabs}
\usepackage{tikz}
\usepackage{graphicx}
\usetikzlibrary{arrows, positioning, fit, shapes.geometric}
\usepackage{pgfplots}
\pgfplotsset{compat=newest}
\usepgfplotslibrary{groupplots}

\usepackage[backend=biber,bibencoding=utf8,giveninits=true,maxbibnames=5,autolang=other,isbn=false]{biblatex}
\newcommand{\doctitle}{Centrality of nodes in Federated Byzantine Agreement Systems}
\newcommand{\docauthors}{André Gaul\footnotemark[1]\; and Jörg Liesen\footnotemark[1]}
\newcommand{\docdate}{\today}

\usepackage{hyperref} 
\hypersetup{
  draft=false,
  pdftitle={\doctitle},
  pdfauthor={Andr\'e Gaul, Jörg Liesen},
  pdflang={en},
  colorlinks=true,
  linkcolor=black,
  citecolor=black,
  filecolor=black,
  urlcolor=black
}

\theoremstyle{definition}
\newtheorem{thm}{Theorem}[section] 

\newtheorem{definition}[thm]{Definition}

\newtheorem{ex}[thm]{Example}

\title{\doctitle}
\author{\docauthors}
\date{\docdate}

\providecommand{\keywords}[1]
{{\small	
\textbf{\textbf{Keywords --}} #1
}}

\begin{document}

\renewcommand{\thefootnote}{\fnsymbol{footnote}}
\footnotetext[1]{TU Berlin, Institute of Mathematics, Stra{\ss}e des 17. Juni 136, 10623 Berlin, Germany, {\tt andre@gaul.io}, {\tt liesen@math.tu-berlin.de}} 
\maketitle

\renewcommand{\thefootnote}{\arabic{footnote}}

\begin{abstract}
The \emph{federated Byzantine agreement system (FBAS)} is a consensus model introduced by Mazi\`eres in 2016~\cite{Maz16} where the participating nodes conceptually form a network, with links between them being established by each node individually and thus in a decentralized way. An important question is whether these decentralized decisions lead to an overall decentralized network. The level of (de-)centralization in a network can be assessed using centrality measures.  In this paper we consider three different approaches for obtaining centrality measures for the nodes in an FBAS. Two of them are based on adapting well-known measures based on graphs and hypergraphs to the FBAS context. Since the network structure of an FBAS can be more complex than (usual) graphs or hypergraphs, we also develop a new, problem-adapted centrality measure. This new measure is based on the intactness of nodes, which is an important ingredient of the FBAS model. We illustrate advantages and disadvantages of the three approaches on several computed examples. We have implemented all centrality measures and performed all computations in the Python package \emph{Stellar Observatory}\/\footnote{\url{https://github.com/andrenarchy/stellar-observatory}}.
\end{abstract}


\section{Introduction}

Fault-tolerant agreement systems have been analyzed and used in the context of computer networks (at least) since the early 1980s~\cite{Pea80}. The goal of such systems is to make the network resilient against the failure of one or more of its nodes or, as stated by Lamport, Shostak and Pease in their classical paper on the Byzantine generals problem~\cite{LamShoPea82}, ``to ensure that the loyal generals will reach agreement''. Nowadays, the term ``agreement'' is often replaced by ``consensus'', and recent years have seen a dynamic development of different \emph{consensus models}, particularly in the context of blockchains and digital currencies; see, e.g.,~\cite{Bano2019} for a survey. One of these models, the \emph{federated Byzantine agreement system (FBAS)}, was introduced by Mazi\`eres in 2016~\cite{Maz16}. It is the mathematical basis of the \emph{Stellar Consensus Protocol}\/\footnote{\url{https://www.stellar.org}}, and our focus in this paper. 

The nodes of an FBAS conceptually form a network, where the links between the nodes are established by trust decisions that are made by each node individually, and hence in a decentralized way. A subset of nodes that a given node ``trusts'' is called a \emph{quorum slice}, and each node may have several such slices. A \emph{quorum} of the FBAS is a subset of nodes that contains at least one quorum slice for each of its elements. 

The general idea behind this model, and possibly its applicability to real-world situations beyond the Stellar Consensus Protocol, can be readily understood when considering the following example: Think of a person who has several different groups of people as friends (e.g. at work, in a sports club, neighbors, or family members). Suppose that this person agrees with some statement when an entire group of friends agrees. Then each of these groups would be a quorum slice in the FBAS model. A quorum then is a group of people, so that for each person in the quorum at least one group of their friends is contained in the quorum. This could be any of the different groups of friends, but for each member of the quorum it needs to be an entire group. A quorum can thus be interpreted as a subset of the nodes (or people) that ``trusts itself''. A simple (abstract) illustration of an FBAS is given in Example~\ref{ex:FBAS_fig7} below. Further examples are given throughout the paper; see in particular Section~\ref{sec:examples}. 

Consensus in the network means that two contradictory statements should not be ratified at the same time, or that all functioning (or well-behaved) nodes should agree on the same values; see the quote from~\cite{LamShoPea82} above. In the FBAS model this is guaranteed when all quorums are pairwise intersecting. The question of quorum intersection, and several other FBAS concepts and extensions lead to challenging mathematical and computational problems, with some of them still under investigation. In addition to Mazi\`eres' paper~\cite{Maz16}, an introductory analysis and FBAS algorithms can be found in~\cite{GauKLS19}. Numerous further recent publications on different aspects show the growing interest in understanding the complex mathematics behind the FBAS consensus model; see, e.g.,~\cite{ChiC19,FloHS20,GarG18,GarS19,Lac19,LokLosMaz19,SonDan19}. 

An essential question about the FBAS is whether the individual and hence decentralized trust decisions of the nodes lead to a network that can also be characterized as decentralized. The level of (de-)centralization in networks can be assessed using centrality measures, and in this paper we study such measures for the FBAS networks. Centrality measures are well established for networks that are represented by graphs or hypergraphs. For graphs they are typically based on the adjacency matrix, with important examples given by the eigenvector and the subgraph centrality measures; see, e.g.,~\cite[Chapter~7]{Est11}, or~\cite{EstHig10} for a brief overview. We adapt both these measures to the FBAS context and apply them to the associated \emph{trust graph}. In addition, since the the nodes of an FBAS and the set of the quorums form a hypergraph, we can also adapt hypergraph-based centrality measures from the literature to the FBAS context. Here one finds similar ideas based on the incidence or the adjacency matrix, which also lead to an eigenvector as well as a subhypergraph centrality.  

In general, however, the FBAS model cannot be fully represented by graphs or hypergraphs, so that the derivation of additional, more problem-adapted centrality measures is of interest. In this paper we introduce a centrality measure that involves the \emph{intactness} of the nodes, which is an important ingredient of the FBAS consensus model when investigating the effect of ill-behaved nodes, e.g.,~failed, misconfigured or malicious nodes. Thus, besides giving new insights into this particular model, our paper can be seen as a case study of how to measure node centrality in networks that are more complex than what can be represented as (usual) graphs or even hypergraphs.   

The only other discussion of node centrality in the FBAS consensus model we are aware of is given in the paper~\cite{KimKK19}, which is devoted to an analysis of the security of the Stellar network in its setup of January 2019. The authors compute the PageRank~\cite{BriPag98} for the trust graph of this network, and they introduce the NodeRank measure. The latter is based on PageRank, and additionally takes threshold values of quorum slices into account. This feature is not explicitly a part of the general FBAS definition, but rather is specific to the definition of the quorum slices in the Stellar network; see~\cite[Section~2.3]{GauKLS19} for a mathematical definition of the threshold concept. The computational results in~\cite{KimKK19} show that PageRank and NodeRank yield similar centrality values for the nodes in the Stellar network of January 2019. The conclusion drawn from~\cite[Figure~3]{KimKK19} is that this network was highly centralized, with only a few nodes dominating all others. An analysis of the effect of highly central nodes (which are trusted by all other nodes) in the Stellar network of January 2019 using the Penrose-Banzhaf index can be found in~\cite{BraGrodeH19}, where the authors call the FBAS a \emph{Byzantine trust network}, and point out the relation between the FBAS and some models from game theory. The different centrality measures in our paper confirm the observations in~\cite{KimKK19} about the centralization of the Stellar network in 2019; see Example~\ref{ex:stellar-old} below. In addition, the measures show that the current setup of the network (of December 2020) is more decentralized; see Examples~\ref{ex:stellar-new}--\ref{ex:stellar-new-3}. Our focus in this paper is however not on a certain setup of the Stellar network, but, as mentioned above, on mathematically derived and computationally tested centrality measures for the FBAS consensus model in general.     

The paper is organized as follows. In Section~\ref{sec:FBAS} we briefly summarize the most important definitions and ideas in the context of the FBAS consensus model. We then derive the different centrality measures in Sections~\ref{sec:trust-graph-centrality}--\ref{sec:intactness-centrality}. These approaches are illustrated and evaluated with computed examples in Section~\ref{sec:examples}.

\section{Federated Byzantine agreement systems}\label{sec:FBAS}

Let us briefly recall the most important definitions in the context of the FBAS. These were originally given by Mazi\`eres in~\cite{Maz16}, and here we take the formulations from~\cite{GauKLS19}.

\begin{definition}\label{def:FBAS}
A \emph{federated Byzantine agreement system (FBAS)} is a pair $(V,S)$ consisting of a finite set of nodes $V$ and a \emph{quorum function} $S:V\to {\mathcal P}({\mathcal P}(V)) \setminus \{\emptyset\}$, where for each $v\in V$ and $s\in S(v)$ we require that $v\in s$. Each set $s\in S(v)$ is called a \emph{quorum slice} of the node $v$. \\
A nonempty set of nodes $Q\subseteq V$ is called a \emph{quorum} in $(V,S)$ if for each $v\in Q$ there exists a quorum slice $s\in S(v)$ with $s\subseteq Q$.
\end{definition}

Note that trivially $Q=V$ is a quorum in $(V,S)$, so that each FBAS has at least one quorum. Moreover, it is easy to see that if $Q_1,Q_2$ are quorums, then $Q_1\cup Q_2$ is a quorum as well. 

The following example was also considered in~\cite{Maz16} and~\cite{GauKLS19}.

\begin{ex}\label{ex:FBAS_fig7}
    Consider the FBAS $(V,S)$ defined by $V=\{1,2,3,4,5,6,7\}$ and $S$ with 
    \begin{align*}
        S(i)&=\{\{1,2,3,7\}\},\; i=1,2,3,\qquad S(i)=\{\{4,5,6,7\}\},\;i=4,5,6,
    \end{align*}
    and $S(7)=\{\{7\}\}$. This FBAS has the four quorums
    \begin{align*}
        \{1,2,3,7\},\quad\{4,5,6,7\},\quad\{7\},\quad V,
    \end{align*}
    which all intersect in the node $7$.
\end{ex}

We point out that in a paper on Stellar as a global payment network, the authors define a quorum as ``a nonempty set of nodes encompassing at least one quorum slice of each \emph{non-faulty} member''~\cite[p.~4]{LokLosMaz19}. The differences between this definition and Definition~\ref{def:FBAS} are discussed in detail in~\cite[Section~4.2]{GauKLS19}.

\medskip
In a nutshell (see~\cite[Section~5.1]{Maz16} for details), a quorum $Q$ ratifies a statement if every node $v\in Q$ asserts that this statement is true. Consensus means that two contradicting statements should not be ratified at the same time. Thus, the quorums of the FBAS should pairwise intersect, which leads to the following definition.

\begin{definition}
    An FBAS $(V,S)$ has \emph{quorum intersection} if any two of its quorums have a nonempty intersection.
\end{definition}

The decision problem whether quorum intersection holds is NP-complete. A proof of this fact and algorithms for quorum enumeration as well as checking quorum intersection are described in~\cite[Section~3]{GauKLS19}. The algorithms are implemented in the Python package \emph{Stellar Observatory}. 

\section{Trust graph-based centrality}\label{sec:trust-graph-centrality}

In this section we will extend the centrality measures for graphs from~\cite{Bon72} and~\cite{EstR05} to the FBAS context. 

\subsection{The trust graph of an FBAS}

One of the easiest ways to get an overview of an FBAS is to look at its trust graph as defined in~\cite[Definition~2.13]{GauKLS19}:

\begin{definition}
    The \emph{trust graph} of the FBAS $(V,S)$ is the directed graph $G=(V,E)$, where for every $u,v\in V$ we have $(u,v)\in E$ if  
    $v\in s$ for some $s\in S(u)$.
\end{definition}

\begin{ex}
    \begin{figure}
        \centering
        \includegraphics[width=0.6\columnwidth]{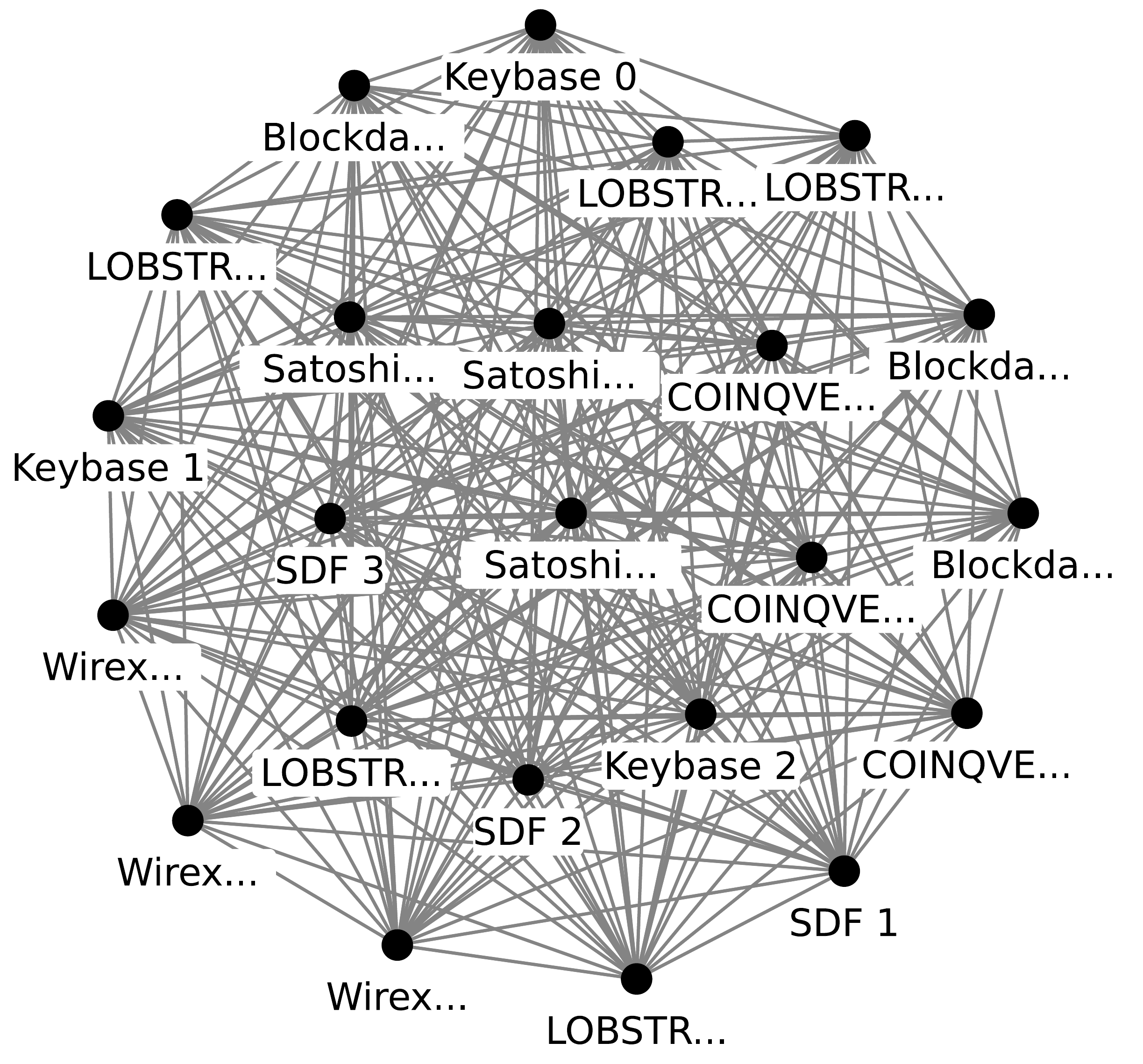}
        \caption{Trust graph of the greatest SCC of the Stellar network in December 2020.}\label{fig:stellarbeat}
    \end{figure}
    Figure~\ref{fig:stellarbeat} is a visualization of the trust graph formed by the greatest SCC of the Stellar network, which consists of $7$ organizations running a total of $23$ nodes\/\footnote{Taken from \url{https://www.stellarbeat.io} on December 3, 2020.}.
    The FBAS formed by these nodes has 1.900.544 quorums, with the smallest quorums containing 10 nodes. Neither this information, nor the intricate definition of the quorum slices can be seen in the trust graph, though.  
\end{ex}

Clearly, the trust graph cannot capture the entire behavior of an FBAS since it does not represent any information about the individual quorum slices. Nevertheless, considering the (significantly) simpler structure of the trust graph instead of the full FBAS can be beneficial. An example is given by the preprocessing step in the quorum intersection algorithm presented in~\cite[Section~3.3]{GauKLS19}. 

Since the trust graph $G=(V,E)$ is a directed graph, we can determine the centrality of its nodes using well-known centrality measures from the literature. In order to define these measures, let $G=(V,E)$ be any directed graph with nodes $V=\{v_1,\dots,v_n\}$ and edges $E\subset V\times V$. A \emph{walk} of length $k\geq 1$ in the graph $G$ is an alternating sequence $w_1,e_1,w_2,\dots, e_k,w_k$ of nodes and edges such that $e_i=(w_i,w_{i+1})$ holds for $i=1,\dots,k-1$. A \emph{path} in the graph is a walk where the nodes $w_1,\dots,w_{k+1}$ are distinct. The graph $G$ is \emph{strongly connected} when for all pairs $v,w$ of (distinct) nodes there exists a path from $v$ to $w$ in $G$. A nonempty set $C\subseteq V$ is a \emph{strongly connected component} (abbreviated \emph{SCC}) of $G$ if it is strongly connected and no proper superset of $C$ is strongly connected.

A ranking among the SCCs in the FBAS context can be defined as follows; cf.~\cite[Section~2]{GauKLS19}.

\begin{definition}\label{def:SCC-props}
    Let $G=(V,E)$ be the trust graph of the FBAS $(V,S)$, and let $C\subseteq V$ and $D\subseteq V$ be SCCs. We say that $D$ is \emph{reachable} from $C$ if there are $c\in C$ and $d\in D$ such that $d$ is reachable from $c$. We say that an SCC is \emph{maximal} if no other SCC is reachable from it. We say that an SCC is the \emph{greatest} SCC if it is reachable from every other SCC.
\end{definition}

The finiteness of the set $V$ implies that each trust graph has at least one maximal SCC. If there is only one maximal SCC, then it is the greatest SCC. Moreover, the greatest SCC is also maximal. In practical applications, particularly in the Stellar network, we usually assume that the given FBAS has quorum intersection. If this is satisfied, then the greatest SCC of its trust graph exists; see~\cite[Lemma~3.8]{GauKLS19}.

The \emph{adjacency matrix} of the graph $G=(V,E)$ is the $n\times n$ matrix $A=A(G)=[a_{ij}]$, where $a_{ij}=1$ if $(v_i,v_j)\in E$ and $a_{ij}=0$ otherwise. The adjacency matrix is nonnegative and in general nonsymmetric, since we consider directed graphs. 

\begin{ex}\label{ex:FBAS_fig7a}
    The SCCs of the FBAS of Example~\ref{ex:FBAS_fig7} are $\{1,2,3\}$, $\{4,5,6\}$ and $\{7\}$; see Figure~\ref{fig:FBAS_fig7a}. 
    The SCC $\{7\}$ is the only maximal SCC, and it is also the greatest SCC. The adjacency matrix of the trust graph is given by
    $$A=\begin{bmatrix} 
    0 & 1 & 1 & 0 & 0 & 0 & 1\\
    1 & 0 & 1 & 0 & 0 & 0 & 1\\
    1 & 1 & 0 & 0 & 0 & 0 & 1\\
    0 & 0 & 0 & 0 & 1 & 1 & 1\\
    0 & 0 & 0 & 1 & 0 & 1 & 1\\
    0 & 0 & 0 & 1 & 1 & 0 & 1\\
    0 & 0 & 0 & 0 & 0 & 0 & 0
    \end{bmatrix}.$$
    
    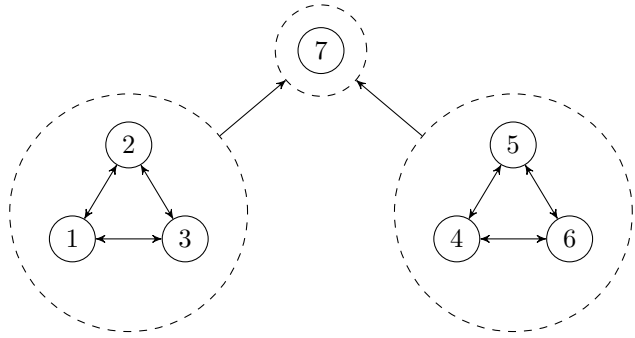
\begin{figure}
        \centering
        \begin{tikzpicture}[auto,
            level distance = 1.25cm,
            node/.style={circle,fill=white,draw},
            edge_style/.style={draw=orange, line width=2, ultra thick},
            bi_dir_e/.style={edge from parent/.style={<->,>=stealth',draw}}]
            
            \node[node] (7) {7};
            \node[circle, draw=black, fit=(7), dashed] (q3) {};
            
            \node[node, below right= 6mm and 19mm of q3] (5) {5}
                child[bi_dir_e]{ node [node] (4) {4}}
                child[bi_dir_e]{ node [node] (6) {6}};
            \draw[<->,>=stealth'] (4) -- (6);
            
            \node[node, below left= 6mm and 19mm of q3] (2) {2}
                child[bi_dir_e]{ node [node] (1) {1}}
                child[bi_dir_e]{ node [node] (3) {3}};
            \draw[<->,>=stealth'] (1) -- (3);
            
            \node[circle, draw=black, fit=(1) (2) (3), dashed, below right=6mm and 10mm of q3] (q2) {};
            \node[circle, draw=black, fit=(4) (5) (6), dashed, below left=6mm and 10mm of q3] (q1) {};
            \draw[->,>=stealth'] (q1) -- (q3);
            \draw[->,>=stealth'] (q2) -- (q3);
        \end{tikzpicture}
        \caption{The strongly connected components of the FBAS from Example~\ref{ex:FBAS_fig7a}.}\label{fig:FBAS_fig7a}  
    \end{figure}
\end{ex}

\begin{ex}\label{ex:FBAS_fig7-mod}
We now add one node to the FBAS of Example~\ref{ex:FBAS_fig7}, which we connect to one of the SCCs, i.e., we consider the FBAS $(V,S)$ with $V=\{1,2,3,4,5,6,7,8\}$ and $S$ with 
    \begin{align*}
        S(i)&=\{\{1,2,3,7\}\},\; i=1,2,3,\qquad S(i)=\{\{4,5,6,7\}\},\;i=4,5,
        \qquad 
    \end{align*}
    $S(6)=\{\{4,5,6,7,8\}\}$, $S(j)=\{\{j\}\}$, $j=7,8$. The FBAS has the quorums
    \begin{align*}
        \{1,2,3,7\},\; \{1,2,3,7,8\},\;
        \{4,5,6,7,8\},\;\{7,8\},\;\{7\},\;\{8\},\; V.
    \end{align*}
    The trust graph and SCCs of this FBAS are shown in Figure~\ref{fig:FBAS_fig7-mod}. There are two maximal SCCs, and thus the greatest SCC does not exist, and we cannot have quorum intersection.
    
    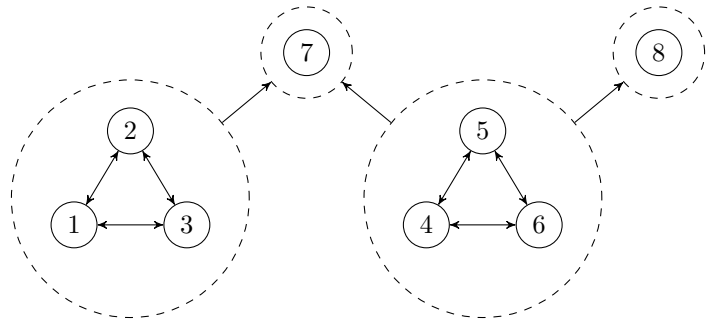
\begin{figure}
        \centering
        \begin{tikzpicture}[auto,
            level distance = 1.25cm,
            node/.style={circle,fill=white,draw},
            edge_style/.style={draw=orange, line width=2, ultra thick},
            bi_dir_e/.style={edge from parent/.style={<->,>=stealth',draw}}]
            
            \node[node] (7) {7};
            \node[circle, draw=black, fit=(7), dashed] (q3) {};
            
            \node[node, below left= 6mm and 19mm of 7] (2) {2}
                child[bi_dir_e]{ node [node] (1) {1}}
                child[bi_dir_e]{ node [node] (3) {3}};
            \draw[<->,>=stealth'] (1) -- (3);
            
            \node[node, below right= 6mm and 19mm of 7] (5) {5}
                child[bi_dir_e]{ node [node] (4) {4}}
                child[bi_dir_e]{ node [node] (6) {6}};
            \draw[<->,>=stealth'] (4) -- (6);
            
            \node[node, above right= 6mm and 19mm of 5] (8) {8};
            \node[circle, draw=black, fit=(8), dashed] (q4) {};
            
            \node[circle, draw=black, fit=(1) (2) (3), dashed, below right=6mm and 10mm of 7] (q2) {};
            \node[circle, draw=black, fit=(4) (5) (6), dashed, below left=6mm and 10mm of 7] (q1) {};
            \draw[->,>=stealth'] (q1) -- (q3);
            \draw[->,>=stealth'] (q2) -- (q3);
            \draw[->,>=stealth'] (q2) -- (q4);
        \end{tikzpicture}
        \caption{The strongly connected components of the FBAS from Example~\ref{ex:FBAS_fig7-mod}.}\label{fig:FBAS_fig7-mod}  
    \end{figure}
    
\end{ex}

\subsection{Trust graph eigenvector centrality}\label{sec:eigenvector-centrality}

The idea of the \emph{eigenvector centrality} of nodes is that the centrality of a node should be proportional to the sum of the centralities of its neighbors, so that a node is central when it has many neighbors, or has neighbors which themselves are central. This idea is often attributed to Bonacich, who presented it in the context of social sciences (and for undirected graphs) in 1972~\cite{Bon72}. It can be found, however, already in Landau's paper on chess rankings as early as 1895~\cite{Landau1895}; see also his paper from 1915~\cite{Landau1915} for a very clear exposition that uses the Perron-Frobenius theory developed in 1907-1909.  

We now transfer this concept to the FBAS context. Let $G=(V,E)$ be the trust graph of an FBAS $(V,S)$. A neighbor of the node $v_i$ in $G$ is a node $v_j$ such that $(v_j,v_i)\in E$. The mathematical idea of the eigenvector centrality then leads to an equation of the form
$$\widehat{c}_{te}(v_i)=\frac{1}{\lambda}\sum_{(v_j,v_i)\in E}\widehat{c}_{te}(v_j),\quad i=1,\dots,n,$$
for some $\lambda>0$. Equivalently, 
$$\widehat{c}_{te}^T A =\lambda \widehat{c}_{te}^T,\quad \widehat{c}_{te}^T=[\widehat{c}_{te}(v_1),\dots,\widehat{c}_{te}(v_n)],$$
where $A=A(G)$ is the adjacency matrix of the graph. Hence the vector $\widehat{c}_{te}$ is a left eigenvector of the matrix $A$ corresponding to the eigenvalue $\lambda$, which we determine next.

If the trust graph is strongly connected, then its adjacency matrix is irreducible (and vice versa), and in this case the Perron-Frobenius Theorem guarantees that $A$ has a simple maximal positive eigenvalue $\lambda$ with a corresponding positive (left) eigenvector $x$; see, e.g.,~\cite[Theorem~8.4.4]{HorJoh13}. Using the normalized entries of this eigenvector we define the \emph{trust graph eigenvector centrality} of the nodes as
\begin{equation}\label{eqn:cte}
c_{te}(v_i):=x_i/(\max_{j} x_j),\quad i=1,\dots,n.
\end{equation}
Note that this centrality measure can be seen as a more sophisticated concept than the \emph{degree centrality}, which only counts the (inbound) degree of a node, and which we do not consider here. 

If an FBAS has multiple SCCs (as in Examples~\ref{ex:FBAS_fig7a} and~\ref{ex:FBAS_fig7-mod}), then the adjacency matrix of its trust graph is reducible, and the trust graph eigenvector centrality is not well defined for the entire FBAS. In this case we could apply the trust graph eigenvector centrality to the subgraph of each (maximal) SCC seperately. In particular, when the FBAS has quorum intersection the greatest SCC of its trust graph exists, and the eigenvector centralities for the nodes of the corresponding subgraph are well defined.

\subsection{Trust graph subgraph centrality}\label{sec:subgraph-centrality}

The idea of the \emph{subgraph centrality} of nodes in a graph or network, first defined in~\cite{EstR05}, is that a node is central when through this node a large part of the graph is reachable. 

We now transfer this concept to the FBAS context. Let $(V,S)$ be an FBAS, and let $G=(V,E)$ be its trust graph. 
The number of walks of length $k\geq 1$ from node $v_i$ to node $v_j$ in $G$ is given by $(A^k)_{ij}$, i.e., the $(i,j)$-entry in the $k$th power of the adjacency matrix $A$.  Since longer walks are ``less important'' than shorter walks, higher powers of the adjacency matrix should be scaled down. The authors of~\cite{EstR05} suggested taking $1/k!$ as the scaling factor for $A^k$, and for the trust graph $G$ this gives the \emph{trust graph subgraph centrality}
\begin{align}\label{eqn:cts}
\widehat{c}_{ts}(v_i) &:=\sum_{k=0}^\infty \frac{1}{k!}(A^k)_{ii}=(\exp(A))_{ii},\quad\mbox{and}\nonumber\\ 
c_{ts}(v_i)&:=\widehat{c}_{ts}(v_i)/(\max_j \widehat{c}_{ts}(v_j)),
\quad i=1,\dots,n.
\end{align}
Note that the subgraph centrality is defined in~\cite{EstR05} (and other related publications, e.g.~\cite{EstHig10}) only for undirected graphs and hence symmetric adjacency matrices. But since the interpretation of the entries of the powers $A^k$ as walks of length $k$ is valid also for undirected graphs, and  $\exp(A)$ is defined for any (square) real or complex matrix $A$, the trust graph subgraph centrality is well defined for any given FBAS, even without quorum intersection.

\section{Hypergraph-based centrality}\label{sec:hypergraph-based}

In this section we will extend two well-known centrality measures for hypergraphs to the FBAS context.

Let $(V,S)$ with $V=\{v_1,\dots,v_n\}$ be an FBAS, and let $Q=\{Q_1,\dots,Q_m\}$ be the corresponding set of the quorums. Then the pair $(V,Q)$ is a \emph{hypergraph}, where each set $Q_j$ is called a \emph{hyperedge}; see~\cite{Ber89} for a classical introduction into the area of hypergraphs. 
The hypergraph $(V,Q)$ with $n$ nodes and $m$ hyperedges can be represented by an $n\times m$ \emph{incidence matrix} $M=[m_{ij}]$, where $m_{ij}=1$ if $v_i\in Q_j$, and $m_{ij}=0$ otherwise. Thus, the rows of the incidence matrix represent the nodes, and the columns represent the quorums of the given FBAS. The $n\times n$ \emph{adjacency matrix} $A=[a_{ij}]$ of the hypergraph is defined by
$$a_{ij}=\left|\{Q_k \mid \{v_i,v_j\}\subseteq Q_k\}\right|,\quad i\neq j,$$
and $a_{ii}=0$ for $i=1,\dots,n$. Thus, the off-diagonal entry $a_{ij}$ is given by the number of hyperedges that contain both nodes $v_i$ and $v_j$. Note that $A=MM^T-{\rm diag}(MM^T)$.

We first describe an eigenvector-based centrality measure for the hypergraph $(V,Q)$, which follows the main ideas from~\cite{BonHolJoh04} (see also~\cite{BorEve97}). Analogously to the development in Section~\ref{sec:eigenvector-centrality}, a node in the FBAS hypergraph $(V,Q)$ may be considered central if it is contained in many quorums or in central quorums, which (similar to \eqref{eqn:cte}) leads to an equation of the form 
\begin{equation}\label{eqn:qe1}
\widehat{c}_{qe} = \frac{1}{\lambda} My,\quad \widehat{c}_{qe}^T=[\widehat{c}_{qe}(v_1),\dots,\widehat{c}_{qe}(v_n)],
\end{equation}
for some $\lambda>0$, and where the vectors $\widehat{c}_{qe}\in {\mathbb R}^{n}$ and $y\in {\mathbb R}^m$ contain the centrality scores of the nodes and the quorums, respectively. On the other hand, a quorum may be considered central if it contains central nodes, which yields the equation
\begin{equation}\label{eqn:qe2}
y = \frac{1}{\lambda} M^T \widehat{c}_{qe}.    
\end{equation}
Using \eqref{eqn:qe2} in \eqref{eqn:qe1} yields 
\begin{equation}\label{eqn:qe3}
MM^T \widehat{c}_{qe}=\lambda^2 \widehat{c}_{qe}.    
\end{equation}
Thus, the vector $\widehat{c}_{qe}$ is an eigenvector of the nonnegative and symmetric positive (semi-)definite matrix $MM^T$. Note that \eqref{eqn:qe1} and \eqref{eqn:qe2} also yield the equation $M^TMy=\lambda^2 y$, which can be used to define the centrality of the quorums. 
 
Since $V$ is a quorum in any given FBAS $(V,S)$, the corresponding incidence matrix $M$ contains a column of the form $[1,\dots,1]^T$. Thus, $MM^T$ is irreducible, and there exists a simple maximal positive eigenvalue $\lambda^2$ with a corresponding positive eigenvector $x$. The normalized components of this eigenvector give the \emph{quorum eigenvector centrality} of the FBAS $(V,S)$, i.e.,
\begin{equation}\label{eqn:cqe}
c_{qe}(v_i):=x_i/(\max_j x_j),\quad i=1,\dots,n.   
\end{equation}
We stress that unlike the trust graph eigenvector centrality, the quorum eigenvector centrality is defined for any given FBAS, even without quorum intersection. 

Following the ideas in~\cite{EstRod06}, we can also transfer the idea of the trust graph subgraph centrality \eqref{eqn:cts} to hypergraphs. A \emph{walk} from node $w_1\in V$ to node $w_{k+1}\in V$ in the hypergraph $(V,Q)$ is an alternating sequence $w_1,Q_1,w_2,\dots,$ $Q_k,w_{k+1}$, where $Q_i\in Q$ and $\{w_i,w_{i+1}\}\subseteq Q_i$ for $i=1,\dots,k$. 
%
Similar to graphs, the number of walks of length $k\geq 1$ from node $v_i$ to $v_j$ in the hypergraph $(V,Q)$ is given by $(A^k)_{ij}$, where $A$ is the adjacency matrix. Analogously to \eqref{eqn:cts}, the \emph{quorum subhypergraph centrality} of the node $v_i$ is now defined by
\begin{align}\label{eqn:cqs}
\widehat{c}_{qs}(v_i) &:=(\exp(A))_{ii},\quad\mbox{and}\nonumber\\
c_{qs}(v_i)&:=\widehat{c}_{qs}(v_i)/(\max_j \widehat{c}_{qs}(v_j)),\quad i=1,\dots,n.    
\end{align}
Similar to the trust graph subgraph centrality \eqref{eqn:cts}, the quorum subhypergraph centrality \eqref{eqn:cqs} is defined for any given FBAS, even without quorum intersection.

When the FBAS has many quorums and many nodes in their intersections, the adjacency matrix $A\in {\mathbb R}^{n\times n}$ potentially has very large entries, so that a direct application of functions that compute $\exp(A)$ (such as {\tt expm} in MATLAB or Python's {\tt scipy.linalg.expm}) may lead to overflows. Using the identity $\exp(A)=\exp(A/s)^s$, which holds for all $s\in {\mathbb C}$, for scaling down $A$ before computing the matrix exponential would require a very large value of~$s$. This in turn would result in taking a very large matrix power and also lead to overflows.

In order to avoid this situation we can use that the real symmetric matrix $A$ has an orthogonal eigendecomposition of the form $A=U\Lambda U^T$ with $U^TU=I$, and $\Lambda={\rm diag}(\lambda_1,\dots,\lambda_n)$, so that
\begin{equation}\label{eqn:expA}
(\exp(A))_{ii}=\sum_{j=1}^n e^{\lambda_j}u_{ij}^2,\quad i=1,\dots,n,
\end{equation}
which is the expression used to define the subhypergraph centrality in~\cite[p.~585]{EstRod06}. Assume that the eigenvalues are ordered increasingly, $\lambda_1\leq\cdots\leq\lambda_n$, then \eqref{eqn:expA} yields
$$(\exp(A))_{ii}=e^{\lambda_n} \sum_{j=1}^n e^{\lambda_j-\lambda_n}u_{ij}^2,\quad i=1,\dots,n,$$
where all terms of the sum are bounded by $1$ in modulus. Instead of computing \eqref{eqn:cqs} we thus compute 
$$\widetilde{c}_{qs}(v_i):=\sum_{j=1}^n e^{\lambda_j-\lambda_n}u_{ij}^2,\quad i=1,\dots,n,$$
and obtain $c_{qs}(v_i)=\widetilde{c}_{qs}(v_i)/(\max_{j} \widetilde{c}_{qs}(v_j))$ for $i=1,\dots,n$.

\section{Intactness-based centrality}\label{sec:intactness-centrality}

In this section we will derive a centrality measure based on the \emph{intactness of nodes}, a concept originally defined in~\cite{Maz16}; see also~\cite[Section~4]{GauKLS19}. 

A node $v\in V$ can be either \emph{well-behaved} or \emph{ill-behaved}. This local property is independent of the quorum function $S$. A node can be ill-behaved for several reasons, for example due to a failure or misconfiguration, it might be compromised or it might have malicious intentions. An important question is whether ill-behaved nodes can negatively impact other (well-behaved) nodes. If from the viewpoint of a given (well-behaved) node all ill-behaved nodes are in  ``unimportant'' or \emph{dispensible} parts of the FBAS, then everything is fine, and the node is \emph{intact}. If, on the other hand, some ill-behaved nodes are contained in ``important'' or \emph{non-dispensible} parts of the FBAS, then the given (well-behaved) node may no longer function well and becomes \emph{befouled}. In order to formalize these intuitive ideas we need the following definition.     

\begin{definition}\label{def:DSet}
    Let $(V,S)$ be an FBAS and let $D\subseteq V$. Then the FBAS 
    $(V,S)^D=(V\setminus D,S^D)$ is defined by
    $$S^D(v)=\{s\setminus D\mid s\in S(v)\}~\text{for all}~v\in V\setminus D.$$
    A set $D\subseteq V$ is a \emph{dispensible set} or \emph{DSet}, when
    \begin{itemize}            
    \item  $(V,S)^D$ has quorum intersection, and
    \item either $V\setminus D$ is a quorum in $(V,S)$ or $D=V$.
    \end{itemize}
\end{definition}

By this definition, $D=V$ is always a DSet. Moreover, if the FBAS $(V,S)$ has quorum intersection, then $D=\emptyset$ is a DSet. An important observation was made in~\cite[Theorem~1]{Maz16}: If $Q\subseteq V$ is a quorum in $(V,S)$, and $D\subseteq V$ is any subset with $Q\setminus D\not=\emptyset$, then $Q\setminus D$ is a quorum in $(V,S)^D$.

\begin{ex}\label{ex:dsets}
The DSets of the FBAS in Example~\ref{ex:FBAS_fig7}, which has quorum intersection, are given by $\emptyset$, $\{1,2,3\}$, $\{4,5,6\}$, $V\setminus\{7\}$, and $V$.    

The DSets of the FBAS in Example~\ref{ex:FBAS_fig7-mod}, which does not have quorum intersection, are given by $\{4,5,6,8\}$, $V\setminus \{7\}$, $V\setminus\{8\}$, and $V$.
\end{ex}

Using the definition of dispensible sets we can now define the intact and befouled nodes.

\begin{definition}\label{def:intact}
        Let $I\subseteq V$ be the set of (all) ill-behaved nodes in the FBAS $(V,S)$, and let $v\in V\setminus I$. Then $v$ is called \emph{$I$-intact} if there exists a \emph{single} DSet $D$ which contains \emph{all} ill-behaved nodes but not $v$, i.e., $I\subseteq D\subseteq V\setminus\{v\}$. If $v$ is not $I$-intact, it is called $I$-befouled.
\end{definition}
    
By definition, the problem of deciding whether a given set $D\subseteq V$ is a DSet requires to decide whether $(V,S)^D$ has quorum intersection. Since the quorum intersection decision problem is NP-complete, the DSet decision problem is computationally at least as hard. An algorithm for determining the $I$-intact nodes is derived in~\cite[Section~5]{GauKLS19}, and is implemented in \emph{Stellar Observatory}.

\subsection{Definition of the intactness-based centrality}

In Definition~\ref{def:intact} we have partitioned the set $V$ into the ill-behaved nodes~$I$, and the well-behaved nodes~$V\setminus I$. Moreover, the well-behaved nodes are partitioned into the $I$-intact and the $I$-befouled nodes. The latter set is given by
$$F_I:=\{v\in V\setminus I\mid \mbox{$\nexists$ Dset $D$ with $I\subseteq D\subseteq V\setminus \{v\}$}\}.$$
The idea of the intactness-based centrality is that a node should be central when ill-behavedness leads to befouldness of many other (central) nodes. This idea is similar to the eigenvector centrality, but additionally we now consider a weighting for the possible subsets of ill-behaved nodes. We thus define
\begin{equation}\label{eqn:cie-hat}
\widehat{c}_{ie}(v_i) := \frac{1}{\lambda} \sum_{\substack{I\subseteq V\\ {v_i}\in I}}
w_{I}\sum_{v_j\in F_I} \widehat{c}_{ie}(v_j),\quad i=1,\dots,n,
\end{equation}
for some $\lambda>0$ and weights $w_I>0$. Equation \eqref{eqn:cie-hat} is the eigenvalue problem $A \widehat{c}_{ie} = \lambda \widehat{c}_{ie}$ with $A=[a_{ij}]$ and 
\begin{equation}\label{eqn:aij-intact}
a_{ij} = \sum_{\substack{ I\subseteq V\\ {v_i}\in I}} w_I \cdot | \{v_j\} \cap F_I|\geq 0.
\end{equation}
Unless we have a specific knowledge about the ill-behavedness of certain nodes, the weights should be decreasing with $|I|$, for example $w_{I}=1/|I|$ or $w_{I}=2^{-|I|}$. Choosing all weights positive guarantees that the matrix $A$ is nonzero, except for the trivial case that
$F_I=\emptyset$ for all nonempty subsets $I\subseteq V$ of ill-behaved nodes.  

If the matrix $A$ is irreducible and thus has a (simple) largest eigenvalue $\lambda>0$ with a corresponding nonnegative eigenvector $x$, we can define the \emph{intactness eigenvector centrality} by the normalized entries of $x$, i.e. $c_{ie}(v_i):= x_i/(\max_j x_j)$, $i=1,\dots,n$.

For a general FBAS (even with quorum intersection), it is not immediately clear whether the matrix $A$ with the entries given in \eqref{eqn:aij-intact} is irreducible. In order to overcome this difficulty and guarantee a well defined centrality measure for any FBAS, we assume that every node $v_i$ has some (small) \emph{base centrality} $b_i\geq 0$, which is added to the right hand side in \eqref{eqn:cie-hat}. We thus define 
\begin{equation}\label{eqn:cil-hat}
\widehat{c}_{il}(v_i) := \Big(\mu \sum_{\substack{I\subseteq V\\ {v_i}\in I}}
w_{I}\sum_{v_j\in F_I} \widehat{c}_{il}(v_j)\Big)+b_i,\quad i=1,\dots,n,
\end{equation}
for some $\mu>0$ and weights $w_I\geq 0$. This gives the linear algebraic system
$(I-\mu A)\widehat{c}_{il}=b,$
where $b=[b_1,\dots,b_n]^T$. This reminds of the linear algebraic system for the \emph{Katz centrality}~\cite{Kat53}. We assume that at least one of the base centralities $b_i$ is larger than zero, so that $b\neq 0$. 

If we choose the parameter $\mu$ with $0<\mu<\|A\|_2^{-1}$ (or $\mu =1$ in the trivial case $A=0$), then $I-\mu A$ is nonsingular and its inverse is given by $(I-\mu A)^{-1}=\sum_{k=0}^\infty (\mu A)^k$. Since $A$ is nonnegative, this also shows that $(I-\mu A)^{-1}$ is nonnegative, and hence we obtain a uniquely determined nonnegative solution $\widehat{c}_{il}=(I-\mu A)^{-1}b$. Using the normalized entries of $\widehat{c}_{il}$ we define the \emph{intactness linear system centrality}, i.e.,
\begin{equation}\label{eqn:cil}
c_{il}(v_i):=\widehat{c}_{il}(v_i)/(\max_j \widehat{c}_{il}(v_j)), \quad i=1,\dots,n. 
\end{equation}

Note that for $\mu\rightarrow 0$ we have $\widehat{c}_{il}\rightarrow b$ in \eqref{eqn:cil-hat}. 
If we increase $\mu$, then the influence of the base centralities on $\widehat{c}_{il}$ (and hence on $c_{il}$) decreases. Choosing $\mu$ too close to the upper bound can lead to a highly ill-conditioned matrix $I-\mu A$, and hence a numerically ill-determined solution of the linear algebraic system. In our experiments we usually choose $\mu=0.5\cdot \|A\|_2^{-1}$.      

\subsection{A strategy that overcomes redundancies}

The intactness linear system centrality $c_{il}$ takes into account all possible subsets $I\subseteq V$ as sets of ill-behaved nodes, and all resulting $I$-befouled nodes $F_I$. This strategy can lead to redundancies, which can be seen from the following example. 

\begin{ex}
Consider the FBAS from Example~\ref{ex:FBAS_fig7}; see Example~\ref{ex:dsets} for the corresponding DSets. If $I=\{1\}$, then the $I$-befouled nodes are given by $F_I=\{2,3\}$. Thus the node $1$ gains centrality from the nodes $2$ and $3$. If we now add nodes from the SCC $\{4,5,6\}$ to the set of ill-behaved nodes $I$, then these nodes will also gain centrality from the nodes $2$ and $3$, although the ill-behavedness of nodes in $\{4,5,6\}$ does not have any influence on the nodes $2$ or~$3$. 
\end{ex}

The trust graph $G=(V,E)$ of the FBAS can be used to avoid such redundancies. For a node $v_i$, we define $P(v_i)\subseteq V$ as the set of all nodes that are reachable from $v_i$ in $G$, and we define $C(v_i)\subseteq V$ as the set of all nodes that can reach $v_i$. By observing that an ill-behaved node $v_i$ can only contribute to the befouledness of nodes in $C(v_i)$, we can modify the definition in \eqref{eqn:cil-hat} and obtain 
\begin{equation}\label{eqn:chl-hat}
\widehat{c}_{hl}(v_i) := \Big(\mu \sum_{\substack{I\subseteq P(v_i)\\ {v_i}\in I}}
w_{I}\sum_{v_j\in F_I\cap C(v_i)} \widehat{c}_{hl}(v_j)\Big)+b_i,
\quad i=1,\dots,n,
\end{equation}
where again $\mu>0$, and the weights $w_I\geq 0$ are parameters. As above, this yields a linear algebraic system $(I-\mu B)\widehat{c}_{hl}=b$, and we choose $0<\mu<\|B\|_2^{-1}$ (or $\mu =1$ in the trivial case $B=0$) in order to guarantee a unique and nonnegative solution $\widehat{c}_{hl}$. The normalized entries of $\widehat{c}_{hl}$ give the
\emph{hierarchical intactness linear system centrality}, i.e.,
\begin{equation}\label{eqn:chl}
c_{hl}(v_i):=\widehat{c}_{hl}(v_i)/(\max_j \widehat{c}_{hl}(v_j)), \quad i=1,\dots,n. 
\end{equation}

\section{Computed examples}\label{sec:examples}

In this section we study the six centrality measures listed in Table~\ref{tab:centralities} using computed examples. We only consider FBAS with quorum intersection. An FBAS without quorum intersection is quite meaningless to analyze as a whole. While such FBAS may occur in actual applications such as the Stellar network, they are usually the result of misconfiguration of non-critical nodes, and thus a restricted FBAS with quorum intersection should be considered in such cases. An example is given by the FBAS in Figure~\ref{fig:FBAS_fig7-mod}, where the node~$8$ could be omitted, or one could consider an FBAS restricted to the nodes $\{1,2,3,7\}$ or $\{4,5,6,8\}$.

In computations that require quorums or intact nodes we use the algorithms implemented in \emph{Stellar Observatory}\/\footnote{The Python code for running all examples from this paper is available in the repository \url{https://github.com/andrenarchy/fbas-centrality}.}. The computations for small FBAS examples such as those in this section require only a few seconds on a standard notebook. Because of the NP-completeness of the quorum intersection decision problem, the corresponding computations for large FBAS are in general intractable.

For the intactness-based centralities in \eqref{eqn:cil} and \eqref{eqn:cil} we use $w_I=2^{-|I|}$ as well as $\mu=0.5\cdot\|A\|^{-1}_2$ (for $c_{il}$) and $\mu=0.5\cdot\|B\|^{-1}_2$ (for $c_{hl}$).

\begin{table}
    \centering
    \begin{tabular}{lccc}
        \toprule
        Centrality measure & Symbol & Definition \\
        \midrule
        Trust graph eigenvector & $c_{te}$ & Eq.~\eqref{eqn:cte} \\
        Trust graph subgraph & $c_{ts}$ & Eq.~\eqref{eqn:cts} \\
        Quorum eigenvector & $c_{qe}$ & Eq.~\eqref{eqn:cqe} \\
        Quorum subhypergraph & $c_{qs}$ & Eq.~\eqref{eqn:cqs} \\
        Intactness linear system & $c_{il}$ & Eq.~\eqref{eqn:cil} \\
        Hierarchical intactness linear system & $c_{hl}$ & Eq.~\eqref{eqn:chl} \\
        \bottomrule
    \end{tabular}
    \caption{Overview of the centrality measures.}
    \label{tab:centralities}
\end{table}

\begin{ex}\label{ex:results-1scc}
    Consider the FBAS $(V,S)$ defined by $V=\{1,2,3,4,5\}$ and $S$ with 
    \begin{align*}
        S(1)&=\{\{1,2\}, \{1,3\}, \{1,4\}, \{1,5\}\},\quad S(2)=\{\{1,2\},\{2,3\}\},\\
        S(j)&=\{\{1,j\}\},\quad j=3,4,5.
    \end{align*}
    The quorums of this FBAS are given by the four sets $\{1,j\}$ for $j=2,3,4,5$, and all unions of these sets. Thus, the FBAS has 15 quorums and quorum intersection.
    The 12 DSets of the FBAS are given by
    \begin{align*}
    & \emptyset,\, \{2\},\, \{2, 3\},\, \{2, 3, 4\},\,
    \{2, 3, 5\},\,\{2, 4\},\,\{2, 4, 5\},\,
    \{2, 5\},\,\{4\},\,\{4, 5\},\,\{5\},\, V.
    \end{align*}
    The (nonsymmetric) adjacency matrix of the trust graph $G=(V,E)$ is given by
    $$A=\begin{bmatrix}
    0 & 1 & 1 & 1 & 1\\
    1 & 0 & 1 & 0 & 0\\
    1 & 0 & 0 & 0 & 0\\
    1 & 0 & 0 & 0 & 0\\
    1 & 0 & 0 & 0 & 0
    \end{bmatrix}.$$
    The matrix is irreducible because the FBAS consists of only one SCC. This means that we can compute $c_{te}$ and thus all centralities in Table~\ref{tab:centralities}.
    The computed centralities, rounded to three significant digits, are given in Table~\ref{tab:results-1scc}. 
    \begin{table}
        \centering
        \begin{tabular}{cllllll}
            \toprule
            Node & $c_{te}$   & $c_{ts}$  & $c_{qe}$      & $c_{qs}$     & $c_{il}$     & $c_{hl}$\\
            \midrule
            1    & 1 (1.0)    & 1 (1.0)    & 1 (1.0)      & 1 (1.0)      & 1 (1.0)      & 1 (1.0) \\
            2    & 3 (0.473)  & 2 (0.476)  & 2 (0.584)    & 2 (0.521)    & 4 (0.607)    & 4 (0.607) \\
            3    & 2 (0.696)  & 2 (0.476)  & 2 (0.584)    & 2 (0.521)    & 2 (0.699)    & 2 (0.699) \\
            4    & 3 (0.473)  & 3 (0.424)  & 2 (0.584)    & 2 (0.521)    & 3 (0.647)    & 3 (0.647) \\
            5    & 3 (0.473)  & 3 (0.424)  & 2 (0.584)    & 2 (0.521)    & 3 (0.647)    & 3 (0.647) \\
            \bottomrule
        \end{tabular}
        \caption{Node ranking and computed centralities (cf.~Table~\ref{tab:centralities}) for Example~\ref{ex:results-1scc}. }
        \label{tab:results-1scc}
    \end{table}
    
    As expected, the node $1$ is the most central according to all measures. Beyond the node $1$ there are interesting differences between the measures:
    
    According to the trust graph eigenvector centrality, the node $3$ is more important than the remaining nodes, which is due to the fact that this node has one more neighbor (namely the node $2$). According to the trust graph subgraph centrality, however, the nodes $2$ and $3$ are equally important. This is because the additional neighbor of node $3$ does not lead to more closed walks that start and end at this node in comparison to the number of walks that start and end at node $2$. 

    The quorum eigenvector and quorum subhypergraph centralities both assign the same importance to all nodes except the node $1$. This is not surprising since the quorums are the sets $\{1,j\}$ for $j=2,3,4,5$, and all unions of these sets. 
    
    For the intactness-based centralities we have $c_{il}=c_{hl}$ because the trust graph of the FBAS has only one SCC, and thus $P(v)=C(v)=V$ holds for every $v\in V$. The two intactness-based centralities reveal that node $3$ can befoul more (important) nodes than nodes $4$ and $5$, and that node $2$ has the least power to befoul others. 

    This small example already demonstrates some shortcomings of the trust graph-based and quorum-based centralities, which are both based on a simplified model that can not capture the complex structure of an FBAS. Among the three different approaches, the intactness-based centrality measures clearly yield the most refined information about the importance of nodes in the given FBAS. 
\end{ex}

\begin{ex}\label{ex:results-1scc-mod}
    We now slightly modify the definition of the FBAS from Example~\ref{ex:results-1scc} by changing the quorum slices of node $2$ from $S(2)=\{\{1,2\},\{2,3\}\}$ to
    $$S(2)=\{\{1,2,3\}\}.$$
    This results in the same trust graph, so the corresponding centralities $c_{te}$ and $c_{ts}$ do not change in comparison with those from Example~\ref{ex:results-1scc}. 
    
    The quorums of the FBAS are now given by $\{1,2,3\}$ and $\{1,j\}$, $j=3,4,5$, and all unions of these four sets. This gives 11 quorums, which all contain the node~$1$. The node~$3$ is contained in eight quorums, nodes $4,5$ are both contained in six quorums, and the node~$2$ is only contained in four quorums. This simple observation is also reflected by the ranking of the nodes obtained from the quorum-based centralities $c_{qe}$ and $c_{qs}$; see Table~\ref{tab:results-1scc-mod}.
    
    The DSets are the same as in Example~\ref{ex:results-1scc} and thus the intactness-based centralities $c_{il}$ and $c_{hl}$ yield the same ranking as before.

    \begin{table}
        \centering
        \begin{tabular}{cllllll}
            \toprule
            Node & $c_{te}$   & $c_{ts}$   & $c_{qe}$   & $c_{qs}$     & $c_{il}$     & $c_{hl}$\\
            \midrule
            1    & 1 (1.0)    & 1 (1.0)    & 1 (1.0)    & 1 (1.0)      & 1 (1.0)      & 1 (1.0) \\
            2    & 3 (0.473)  & 2 (0.476)  & 4 (0.431)  & 4 (0.335)    & 4 (0.607)    & 4 (0.607) \\
            3    & 2 (0.696)  & 2 (0.476)  & 2 (0.795)  & 2 (0.792)    & 2 (0.699)    & 2 (0.699) \\
            4    & 3 (0.473)  & 3 (0.424)  & 3 (0.586)  & 3 (0.51)    & 3 (0.647)    & 3 (0.647) \\
            5    & 3 (0.473)  & 3 (0.424)  & 3 (0.586)  & 3 (0.51)    & 3 (0.647)    & 3 (0.647) \\
            \bottomrule
        \end{tabular}
        \caption{Node ranking and computed centralities (cf.~Table~\ref{tab:centralities}) for Example~\ref{ex:results-1scc-mod}. }
        \label{tab:results-1scc-mod}
    \end{table}
\end{ex}

\begin{ex}\label{ex:results-FBAS_fig7}
    We now consider the FBAS from Example~\ref{ex:FBAS_fig7a} which consists of three SCCs; cf. Figure~\ref{fig:FBAS_fig7a}. Because of the multiple SCCs, the trust graph eigenvector centrality $c_{te}$ is not applicable. (At least not to the entire FBAS.) The computed centralities are given in Table~\ref{tab:results-FBAS_fig7}.
    \begin{table}
        \centering
        \begin{tabular}{cllllll}
            \toprule
            Node & $c_{te}$ & $c_{ts}$   & $c_{qe}$   & $c_{qs}$     & $c_{il}$     & $c_{hl}$\\
            \midrule
            1    & --       & 1 (1.0)    & 2 (0.667)  & 2 (0.63)     & 2 (0.835)    & 2 (0.549) \\
            2    & --       & 1 (1.0)    & 2 (0.667)  & 2 (0.63)     & 2 (0.835)    & 2 (0.549) \\
            3    & --       & 1 (1.0)    & 2 (0.667)  & 2 (0.63)     & 2 (0.835)    & 2 (0.549) \\
            4    & --       & 1 (1.0)    & 2 (0.667)  & 2 (0.63)     & 2 (0.835)    & 2 (0.549) \\
            5    & --       & 1 (1.0)    & 2 (0.667)  & 2 (0.63)     & 2 (0.835)    & 2 (0.549) \\
            6    & --       & 1 (1.0)    & 2 (0.667)  & 2 (0.63)     & 2 (0.835)    & 2 (0.549) \\
            7    & --       & 2 (0.369)  & 1 (1.0)    & 1 (1.0)      & 1 (1.0)      & 1 (1.0) \\
            \bottomrule
        \end{tabular}
        \caption{Node ranking and computed centralities (cf.~Table~\ref{tab:centralities}) for Example~\ref{ex:results-FBAS_fig7}. }
        \label{tab:results-FBAS_fig7}
    \end{table}
    
    According to the trust graph subgraph centrality $c_{ts}$, the nodes $\{1,\dots,6\}$ are equally important with centrality 1.0, whereas node $7$ is less important. This result may seem surprising but becomes apparent from the definition of $c_{ts}$, which considers walks in the trust graph. The low importance of the node~$7$ is due to the fact no walks pass through this node in the trust graph pass. Obviously, any useful centrality measure for an FBAS should identify the node~$7$ as the most central node in this example, because this node is the only one that is contained in all quorum intersections. The example therefore illustrates that the quantities measured by the trust graph subgraph centrality may be of little relevance to the actual importance of the nodes in an FBAS. 
    
    While both trust graph-based centrality measures fail in this example, the two other approaches (based on quorums and intactness) indeed show that the node $7$ is the most important one with centrality 1.0, and that the other nodes are less but all equally important. The latter is expected because of the symmetry of the quorum slice definitions.

    In the hierarchical intactness centrality $c_{hl}$ we only consider contributions from nodes that are reachable from a given node. This removes redundancies, which for the nodes $\{1,\dots,6\}$ leads to lower (numerical) values in comparison to $c_{il}$. Thus, according to the centrality measure $c_{hl}$ the nodes $\{1,\dots,6\}$ appear to be even ``less important'' in comparison to the node $7$. 
\end{ex}

\begin{ex}\label{ex:stellar-old}
    We now consider an example modeled after the Stellar network in early 2019. Suppose that the network consists of three organizations which operate a total of 8 nodes nodes, $V=A\cup B\cup C$ with
    $$A=\{a_1,a_2,a_3,a_4\},\quad B=\{b_1,b_2,b_3\},\quad C=\{c\}.$$
    The quorum slices are given as follows:
    \begin{enumerate}
    \item $S(a_j)$ consists of all sets containing $a_j$ and \emph{two} nodes from $A\setminus\{a_j\}$.
    \item $S(b_j)$ consists of all sets containing $b_j$, \emph{one} node from $B\setminus\{b_j\}$ and \emph{three} nodes from $A$.
    \item $S(c)$ consist of all sets containing $c$ and \emph{three} nodes from $A$. 
    \end{enumerate}
    (One could describe this more formally using the definitions introduced in~\cite[Section~2.3]{GauKLS19}.)

    The nodes of each organization form an SCC in the trust graph, and the nodes of organization $A$ form the greatest SCC.
    Because of the multiple SCCs, the trust graph eigenvector centrality $c_{te}$ is not applicable to the entire FBAS.
    The FBAS has 50 quorums and quorum intersection, with every two quorums intersecting in at least two nodes of organization $A$. 
    There are 51 DSets, including $\emptyset$ and $V$.
    The computed centralities are given in Table~\ref{tab:stellar-old}.
    
    All centrality measures yield the expected result that the nodes of organization $A$ are the most central, followed by organization $B$, and organization $C$. The large (numerical) difference between the trust graph subgraph centrality values $c_{ts}$ of the three organizations indicates that the FBAS is (strongly) centralized in organization $A$. A similar observation about the real Stellar network in its configuration of early 2019 was reported in~\cite{KimKK19}, where the authors used PageRank and a modification called NodeRank in order to compute centrality values.
    
    It can be seen from the SCCs that $c$ cannot befoul any other node without ill-behaved nodes in other SCCs, and thus the hierarchical intactness-based centrality $c_{hl}(c)$ is simply given by the (normalized) base centrality. However, any two nodes of organization $B$ can befoul the remaining node of $B$. Thus the sum in~\eqref{eqn:chl-hat} for any node of $B$ is positive, and hence the centralities $c_{hl}$ of the nodes in $B$ are greater than $c_{hl}(c)$.
    
    \begin{table}
        \centering
        \begin{tabular}{cllllll}
            \toprule
            Node & $c_{te}$ & $c_{ts}$   & $c_{qe}$  & $c_{qs}$  & $c_{il}$  & $c_{hl}$\\
            \midrule
            $a_j$ & --       & 1 (1.0)   & 1 (1.0)   & 1 (1.0)   & 1 (1.0)   & 1 (1.0)   \\
            $b_j$ & --       & 2 (0.511) & 2 (0.807) & 2 (0.723) & 2 (0.787) & 2 (0.719) \\
            $c$   & --       & 3 (0.189) & 3 (0.66)  & 3 (0.511) & 3 (0.767) & 3 (0.611) \\
            \bottomrule
        \end{tabular}
        \caption{Node ranking and computed centralities (cf.~Table~\ref{tab:centralities}) for Example~\ref{ex:stellar-old}. }
        \label{tab:stellar-old}
    \end{table}    
    
\end{ex}

\begin{ex}\label{ex:stellar-new}
    We now consider an example modeled after the Stellar network in late 2020. In comparison with Example~\ref{ex:stellar-old}, the idea of the network remodeling is to have more organizations forming the greatest SCC. As in Example~\ref{ex:stellar-old}, suppose that the network is given by $V=A\cup B\cup C$ with $$A=\{a_1,a_2,a_3,a_4\},\quad B=\{b_1,b_2,b_3\},\quad C=\{c\}.$$
    The quorum slices for each node $v\in V$ are now given by all sets containing $v$, three nodes from $A$, two nodes from $B$, and the node $c$. Thus, all quorum slices contain exactly six nodes.

    All nodes form a single SCC in the trust graph, which also is the greatest SCC. The FBAS has 20 quorums and quorum intersection, with every two quorums intersecting in at least four nodes. There are 21 DSets. The computed centralities are given in Table~\ref{tab:stellar-new}. 
    
    The trust graph is complete (i.e., every node points to every other one) and hence the trust graph-based centralities $c_{te}$ and $c_{ts}$ are trivially $1.0$ for all nodes. For the other centralities we observe, in comparison with Example~\ref{ex:stellar-old}, that the computed values are now closer to each other, which indicates that the FBAS is more balanced or decentralized. According to all measures the node $c$ is the most central. For the quorum-based and intactness-based measures this is expected since the node $c$ is contained in every quorum, and in no DSet (except trivially in $V$).  
    
    We analyze the effect of organizations having different numbers of nodes and thresholds in more detail in the next examples.  
    
      \begin{table}
        \centering
        \begin{tabular}{cllllll}
            \toprule
            Node     & $c_{te}$ & $c_{ts}$  & $c_{qe}$  & $c_{qs}$  & $c_{il}$  & $c_{hl}$\\
            \midrule
            $a_j$    & 1 (1.0)  & 1 (1.0)   & 2 (0.806) & 2 (0.7)   & 2 (0.897) & 2 (0.897)   \\
            $b_j$    & 1 (1.0)  & 1 (1.0)   & 3 (0.759) & 3 (0.633) & 3 (0.878) & 3 (0.878) \\
            $c$      & 1 (1.0)  & 1 (1.0)   & 1 (1.0)   & 1 (1.0)   & 1 (1.0)   & 1 (1.0) \\
            \bottomrule
        \end{tabular}
        \caption{Node ranking and computed centralities (cf.~Table~\ref{tab:centralities}) for Example~\ref{ex:stellar-new}. }
        \label{tab:stellar-new}
    \end{table}    
\end{ex}

\begin{ex}\label{ex:stellar-new-2}
    In another example modeled after the Stellar network in late 2020 we consider $V=A\cup B\cup C$ with
    $$A=\{a_1,a_2,a_3\},\quad B=\{b_1,b_2,b_3\},\quad C=\{c_1,c_2,c_3,c_4,c_5\}.$$
    The quorum slices for a node $v\in V$ are now given by all sets containing $v$, and where two of the following three conditions are satisfied:
    \begin{itemize}
        \item[(1)] The set contains \emph{one} node from $A\setminus\{v\}$ if $v\in A$, or \emph{two} nodes from $A$, otherwise.
        \item[(2)] The set contains \emph{one} node from $B\setminus\{v\}$ if $v\in B$, or \emph{two} nodes from $B$, otherwise.
        \item[(3)] The set contains \emph{two} nodes from $C\setminus\{v\}$ if $v\in C$, or \emph{three} nodes from $C$, otherwise.
    \end{itemize}
    
    As in Example~\ref{ex:stellar-new}, all nodes form a single SCC in the trust graph, which also is the greatest SCC. Now the FBAS has 1024 quorums and quorum intersection, and the only DSets are $\emptyset$ and $V$.
    
    Due to the construction of the quorum slices, the nodes of organization $C$ occur in fewer quorums than those of organizations $A$ and $B$ ($608$ vs. $640$). This explains why the nodes of organization $C$ have a lower quorum-based centralities $c_{qe}$ and $q_{qs}$. The large number of quorums and nodes in their intersections in this example leads to large entries in the adjacency matrix $A$, and we have computed the values $q_{qs}$ as described at the end of Section~\ref{sec:hypergraph-based}. (The largest entry of $A$ in this example is 384.)

    As in Example~\ref{ex:stellar-new}, the trust graph is complete, and hence $c_{te}$ and $c_{ts}$ are trivially $1.0$ for all nodes. Moreover, since there are only trivial DSets, all entries of the matrices $A$ and $B$ used to compute the values $c_{il}$ and $c_{hl}$ 
    are nonzero and equal (except for the diagonal, which is zero), and hence the centralities are $1.0$ for all nodes. Consequently, the FBAS is completely decentralized from the viewpoint of the trust graph-based and the intactness-based centralities. However, the FBAS is entirely unresilient against ill-behaved nodes because there are no non-trivial DSets. 

    \begin{table}
        \centering
        \begin{tabular}{cllllll}
            \toprule
            Node     & $c_{te}$ & $c_{ts}$  & $c_{qe}$  & $c_{qs}$  & $c_{il}$  & $c_{hl}$\\
            \midrule
            $a_j$    & 1 (1.0)  & 1 (1.0)   & 1 (1.0)   & 1 (1.0)   & 1 (1.0)   & 1 (1.0)   \\
            $b_j$    & 1 (1.0)  & 1 (1.0)   & 1 (1.0)   & 1 (1.0)   & 1 (1.0)   & 1 (1.0) \\
            $c_j$    & 1 (1.0)  & 1 (1.0)   & 2 (0.958) & 2 (0.931) & 1 (1.0)   & 1 (1.0) \\
            \bottomrule
        \end{tabular}
        \caption{Node ranking and computed centralities (cf.~Table~\ref{tab:centralities}) for Example~\ref{ex:stellar-new-2}. }
        \label{tab:stellar-new-2}
    \end{table}    
\end{ex}

\begin{ex}\label{ex:stellar-new-3}
    We now modify Example~\ref{ex:stellar-new-2} by changing the third condition to:
    \begin{itemize}
    \item[(3')] The set contains \emph{three} nodes from $C\setminus\{v\}$ if $v\in C$, or \emph{four} nodes from $C$, otherwise.
    \end{itemize}
    
    As in Examples~\ref{ex:stellar-new} and~\ref{ex:stellar-new-2}, all nodes form a single SCC in the trust graph, which also is the greatest SCC.
    Although the FBAS appears to be only a minor modification of the one in Example~\ref{ex:stellar-new-2}, it now has 704 quorums (instead of 1024) and 17 DSets (instead of $2$). The DSets are given by all subsets of $C$ containing one or two elements, plus $\emptyset$ and $V$. The FBAS still has quorum intersection.
    
    As in Example~\ref{ex:stellar-new-2}, the trust graph is complete, and hence $c_{te}$ and $c_{ts}$ are trivially $1.0$ for all nodes. 
    Also, the nodes of organization $C$ occur in fewer quorums than those of organizations $A$ and $B$ ($416$ vs. $480$), which explains why the nodes of organization $C$ have lower quorum-based centralities~$c_{qe}$ and $q_{qs}$
    
    Because of the non-trivial DSets we get a more interesting result for $c_{il}$ and $c_{hl}$ than in Example~\ref{ex:stellar-new-2}. From the DSets we see that ill-behavedness of any one or two nodes of organization $C$ has no impact on the intactness of the remaining network. Thus, the nodes from organization $C$ have a lower intactness-based centralities.  

    We stress that the small change of the threshold in condition $(3)$ leads to the existence of more DSets and thus a stronger resilience against ill-behaved nodes in comparison with the FBAS in Example~\ref{ex:stellar-new-2}.

    \begin{table}
        \centering
        \begin{tabular}{cllllll}
            \toprule
            Node     & $c_{te}$ & $c_{ts}$  & $c_{qe}$  & $c_{qs}$  & $c_{il}$  & $c_{hl}$\\
            \midrule
            $a_j$    & 1 (1.0)  & 1 (1.0)   & 1 (1.0)   & 1 (1.0)   & 1 (1.0)   & 1 (1.0)   \\
            $b_j$    & 1 (1.0)  & 1 (1.0)   & 1 (1.0)   & 1 (1.0)   & 1 (1.0)   & 1 (1.0) \\
            $c_j$    & 1 (1.0)  & 1 (1.0)   & 2 (0.891) & 2 (0.828) & 2 (0.965) & 2 (0.965) \\
            \bottomrule
        \end{tabular}
        \caption{Node ranking and computed centralities (cf.~Table~\ref{tab:centralities}) for Example~\ref{ex:stellar-new-3}. }
        \label{tab:stellar-new-3}
    \end{table}    
\end{ex}

\section{Summary and discussion}

The links between the nodes in an FBAS network are established through the individual quorum slices. Small changes of quorum slice definitions can have a significant impact on the properties of the resulting network, which is illustrated by a comparison of Examples~\ref{ex:stellar-new-2} and~\ref{ex:stellar-new-3}. The complex mathematics behind the FBAS model also means that it is not at all obvious whether the decentralized trust decisions lead to an overall decentralized network. In order to assess the level of (de-)centralization in an FBAS, we have derived and analyzed three different types of centrality measures for the nodes. These are based on the trust graph, the hypergraph formed by the quorums, and on the intactness of the nodes of the FBAS. We now briefly summarize our main observations about these measures.

\emph{Trust graph-based:} The trust graph leads to the centrality measures $c_{te}$ (trust graph eigenvector) and $c_{ts}$ (trust graph subgraph). The trust graph is easily formed from the quorum slices of the FBAS, and the two measures can be computed at a (much) lower cost than the other measures  considered in this paper. However, information is usually lost when forming the trust graph, since a directed graph cannot fully model an FBAS. This is illustrated in Example~\ref{ex:stellar-new-2}, where the two trust graph-based centralities yield no distinction between the nodes of the three different organizations (see Table~\ref{tab:stellar-new-2}). The eigenvector centrality $c_{te}$ is defined for the entire FBAS only when is has just a single SCC, and hence was not applicable in our Examples~\ref{ex:FBAS_fig7} and~\ref{ex:stellar-old}. As shown in Example~\ref{ex:FBAS_fig7}, the subgraph centrality $c_{ts}$ is based on information (namely the number of walks in the trust graph) that may be of little relevance for the actual importance of nodes in an FBAS. 

\emph{Quorum-based:} The quorums of an FBAS lead to the centrality measures $c_{qe}$ (quorum eigenvector) and $c_{qs}$ (quorum subhypergraph). Computing the quorums is computationally challenging, as the quorum intersection decision problem is NP-complete. The incidence matrix $M$ fully represents the hypergraph formed by the nodes and the quorums of an FBAS, but the two centrality measures are based on the matrices $MM^T$ or $A=MM^T-{\rm diag}(MM^T)$, and information is lost when forming this matrix. In addition, the quorums may not fully represent the actual structure of an FBAS which is defined by the quorum slices. An illustration is given in Example~\ref{ex:results-1scc}, where the quorum-based centrality measures (unlike all other measures) cannot distinguish between the nodes $\{2,3,4,5\}$ (see Table~\ref{tab:results-1scc}). Note that a small modification of the quorum slices in Example~\ref{ex:results-1scc} leads to a significantly different quorum-based ranking of the nodes in Example~\ref{ex:results-1scc-mod} (see Table~\ref{tab:results-1scc-mod}). 

\emph{Intactness-based:} Using the FBAS intactness concept we have defined the intactness-based centralities $c_{il}$ (linear system) and $c_{hl}$ (hierarchical linear system). The two measures can only differ from each other when the FBAS has more than one SCC, as illustrated in Example~\ref{ex:results-1scc-mod} (see Table~\ref{tab:results-1scc-mod}). In such cases the hierarchical approach avoids redundancies which may artificially inflate the centrality values of some nodes. The intactness-based centralities are more expensive to compute than the quorum-based centralities, since in addition to the quorums the DSets of the FBAS are required. When this information is available, the centralities $c_{il}$ and $c_{hl}$ exist and are computable for any given FBAS. Moreover, as illustrated in Example~\ref{ex:results-1scc}, the intactness-based centralities may give a more refined information than the approaches based on the trust graph or the quorum hypergraph (see Table~\ref{tab:results-1scc}). 


\section*{Acknowledgements}

This work was supported by a grant in the SDF Academic Research Program. We thank Ismail Khoffi and Torsten St{\"u}ber for their input and helpful discussions.

\printbibliography

\end{document}